# Thermal Gravitational Waves from Primordial Black Holes

C Sivaram and Kenath Arun

Indian Institute of Astrophysics, Bangalore

**Abstract:** Thermal gravitational waves can be generated in various sources such as, in the cores of stars, white dwarfs and neutron stars due to the fermion collisions in the dense degenerate Fermi gas [1, 2, 3]. Such high frequency thermal gravitational waves can also be produced during the collisions in a gamma ray burst [3] or during the final stages of the evaporation of primordial black holes [4]. Here we estimate the thermal gravitational waves from primordial black holes and estimate the integrated energy of the gravitational wave emission over the entire volume of the universe and over Hubble time. We also estimate the gravitational wave flux from gamma ray bursts and jets.

## Thermal Gravitational Waves from PBHs

Primordial black holes (PBHs) are hypothetical type of black holes that is formed not by the gravitational collapse of a star but by the extreme densities of matter present during early universe [4]. As these PBHs evaporate through Hawking radiation, part of the energy could be released in the form of thermal gravitational waves. In the case of spontaneous graviton emission, the quadrupole gravitational power is given by: [5]

$$P_{GW} = \frac{G}{c^5} m^2 \omega^6 R^4 \qquad \ldots (1)$$

A typical PBH, that evaporates over the Hubble time has a mass of $\sim 10^{14} g$ and the corresponding Schwarzschild radius $R_S \approx 10^{-13} cm$.

The power due to the gravitational waves can be written as: [6]

$$P_{GW} = \frac{G}{c^5} \left( \frac{m v^2}{t} \right)^2 \qquad \ldots (2)$$

Where v is the typical velocity and *t* is the time scale, of the underlying explosive processes giving rise to the gravitational wave emission.

The term within the bracket in equation (2) corresponds to the power of the explosion, therefore we have:

$$P_{GW} = \frac{G}{c^5} P_{EXP}^2 \qquad \ldots (3)$$

The total energy released in the form of gravitational waves is then:

$$E_{GW} = \left(\frac{G}{c^5} P_{EXP}^2\right) t \qquad \ldots (4)$$

The number of quanta of gravitational waves is then given by: [6]

$$N_{GW} = \frac{E_{GW}}{\hbar/t} = \frac{G}{c^5} \frac{P_{EXP}^2 t^2}{\hbar} = \frac{G}{c^5} \frac{E_{EXP}^2}{\hbar} \qquad \ldots (5)$$

Where, $E_{EXP} = P_{EXP} t$ is the energy of the explosion and $\hbar/t$ is the energy of each gravitational wave quanta.

For the typical PBH of mass $\sim 10^{14} g$, the energy of the explosion is:

$$E_{EXP} = mc^2 \approx 10^{35} ergs \qquad \ldots (6)$$

Using this in equation (5) we get:

$$N_{GW} \approx 10^{38} \qquad \ldots (7)$$

The frequency and the corresponding energy of the gravitational wave quanta are given by:

$$\begin{aligned} \nu &= \frac{1}{t} \\ \varepsilon &= \frac{\hbar}{t} \end{aligned} \qquad \ldots (8)$$

Where $\qquad t = \frac{R_S}{c} \approx 10^{-23} s \qquad \ldots (9)$

Therefore each quanta of gravitational waves has an energy of $\approx 10^{-4} ergs (\sim 100 MeV)$ with a frequency of $\approx 10^{23} Hz$

The total energy associated with the gravitational waves is then given as:

$$E_{GW} = N_{GW} \varepsilon \approx 10^{34} ergs \qquad \ldots (10)$$

We see from equations (6) and (10) that 10% of the energy of PBH explosion is converted to high frequency gravitational waves, with a typical frequency of $\sim 10^{23} Hz$ corresponding to ~100MeV.

According to current estimates, the number density of PBHs in the universe could be $\sim 1/(kpc)^3$ [8]. Then the total energy associated with the gravitational waves from these PBHs over the entire volume of the universe of $(10^{28} cm)^3$ is:

$$E_T \approx (10^{20})(10^{34} ergs) \sim 10^{54} ergs \qquad \ldots (11)$$

This corresponds to an energy density of $\sim 10^{-30} ergs/cc$, and a flux of

$$f_T \approx 10^{-20} ergs/cm^2/s \qquad \ldots (12)$$

in high energy thermal gravitational waves at typical energies of ~100MeV.

If the number density of PBH is $\sim 1/(pc)^3$, then the flux would be

$$f_T \sim 10^{-10} ergs/cm^2/s \qquad \ldots (13)$$

The above discussion pertains to only those PBH's that evaporate over the Hubble time of $10^{10}$ years. The integrated flux from all the PBH's over the entire Hubble time will give the background thermal gravitational flux.

The number of the PBH's as a function of mass can be written as: [8]

$$n_{bh}(m) = n_{bh}(m_0)\left(\frac{m_0}{m}\right)^{-n} \qquad \ldots (14)$$

Where $n_{bh}(m_0)$ is the present number density of PBHs, $m_0$ is the mass of the PBH that evaporate over the Hubble time and n = 3.

The integrated energy of the gravitational wave emission over the entire volume of the universe is given by:

$$E_{total} = \int_{m_1}^{m_2} \frac{m^2 c^4}{\hbar} \frac{c}{H_0} n_{bh}(m_0)\left(\frac{m_0}{m}\right)^{-3} dm \times Volume \qquad \ldots (15)$$

The upper limit of the integral $m_2 = 10^{14} g$ is the mass of the PBH that will evaporate over the Hubble time scale and the lower limit $m_1 \sim 10^{13} g$, below which the flux is too small. Volume $= 2\pi^2 R_H^3$, where $R_H = 10^{28} cm$ is the Hubble radius.

Plugging in the values we get the total integrated energy as:

$$E_{total} \approx 10^{32} ergs \qquad \qquad ... (16)$$

And the total integrated energy flux is given by:

$$f_{total} = \frac{E_{total}}{4\pi R_H^2} \approx 10^{-24} ergs/cm^2/s \qquad \qquad ... (17)$$

The energy associated with each of the thermal gravitational wave quanta is of the order of $100 MeV \approx 10^{-4} ergs$. Therefore the total flux of thermal gravitational wave quanta is:

$$N_{total} \approx 10^{-20} /cm^2/s \qquad \qquad ... (18)$$

If the number density of the PBH's is $\sim 1/(pc)^3$, then the flux will be of the order of $10^{-11}/cm^2/s$

**Gravitational Waves from GRBs**

In the case of GRB, the energy released over a time scale of ~1second, is of the order of $10^{52} ergs$. This gives:

$$N_{GW} \approx 10^{72} \qquad \qquad ... (19)$$

The energy of each of the gravitational wave quanta (at a frequency of ~1 *Hz*) is:

$$\varepsilon = \frac{\hbar}{t} \sim 10^{-27} ergs \qquad \qquad ... (20)$$

The total energy associated with the gravitational waves is then given as:

$$E_{GW} = N_{GW}\varepsilon \approx 10^{45} ergs \qquad \qquad ... (21)$$

In the case of short duration GRBs of $t \approx 10^{-3} s$, the number of gravitational wave quanta remains the same since the energy of the explosion is the same but the energy of each quanta (at a frequency of ~1 *kHz*) will be:

$$\varepsilon = \frac{\hbar}{t} \sim 10^{-24} \, ergs \qquad \ldots (22)$$

And the total energy associated with the gravitational waves is:

$$E_{GW} = N_{GW}\varepsilon \approx 10^{48} \, ergs \qquad \ldots (23)$$

In addition to these GRBs emitting $10^{48} \, ergs$ in *kHz* gravitational wave quanta, it also emits about $10^{51} \, ergs$ in gamma rays and $10^{53} \, ergs$ in neutrinos. For a GRB at 100Mpc the flux corresponding to these emissions are given by:

$$\begin{aligned} f_{GW} &= 10^{-6} \, ergs/cm^2 \\ f_{\gamma} &= 10^{-3} \, ergs/cm^2 \\ f_{\nu} &= 10^{-1} \, ergs/cm^2 \end{aligned} \qquad \ldots (24)$$

**Gravitational Waves from Jets**

Objects as diverse as X-ray binaries, radio galaxies, quasars, and even the galactic centre, are powered by the gravitational energy released when surrounding gas is accreted by the black hole at their cores. Apart from copious radiation, one of the manifestations of this accretion energy release is the production of jets, collimated beams of matter that are expelled from the innermost regions of accretion discs.

The power associated with the jet is given by: [9]

$$P_{jet} = B^2 R^2 c \Delta t \Gamma^2 \qquad \ldots (25)$$

Where B is the magnetic field associated with the jet over a range of R. In the case of jets produced by short lived energetic events like gamma ray bursts, the duration of the event $\Delta t \sim 1s$ and the gamma factor $\Gamma \approx 100$.

For typical values ($B \approx 10^8 \, G$ and $R \approx 10^{10} \, cm$) we have the power associated with the jet as:

$$P_{jet} = 10^{51} \, ergs/s \qquad \ldots (26)$$

And the energy as:

$$E_{jet} = P_{jet} \Delta t = 10^{51} \, ergs \qquad \ldots (27)$$

The number of quanta of gravitational waves from the jet is given by equation (5) as (which is a basic relation, valid for a whole gamut of systems): [7]

$$N_{GW} = \frac{G}{c^5} \frac{E_{jet}^2}{\hbar} \approx 10^{70} \qquad \ldots (28)$$

The energy of each of the gravitational wave quanta (at a frequency of ~1 Hz) is:

$$\varepsilon = \frac{\hbar}{t} \sim 10^{-27} \, ergs \qquad \ldots (29)$$

The total energy associated with the gravitational waves is then given as:

$$E_{GW} = N_{GW} \varepsilon \approx 10^{43} \, ergs \qquad \ldots (30)$$

In the case of galactic jets, the field strength is much lower but it spans over a larger distance. Typically we have $B \approx 10^{-6} G$ and $R \approx 10^{23} cm$. And the typical time scale $\Delta t \sim 10^{13} s$ and the gamma factor ~10.

The energy associated with the jet in this case becomes:

$$E_{jet} \approx 10^{60} \, ergs \qquad \ldots (31)$$

And the number of quanta of gravitational waves from the jet is given by:

$$N_{GW} = \frac{G}{c^5} \frac{E_{jet}^2}{\hbar} \approx 10^{89} \qquad \ldots (32)$$

The total energy associated with the gravitational waves is then given as:

$$E_{GW} = N_{GW} \varepsilon \approx 10^{49} \, ergs \qquad \ldots (33)$$

of very low frequency $\left(\sim 10^{-13} Hz\right)$ gravitational waves.